\documentclass[journal=jctcce,manuscript=article]{achemso}
\setkeys{acs}{usetitle=true}

\usepackage[version=3]{mhchem} 
\usepackage{xr}
\usepackage{amsmath,amssymb}
\usepackage{graphicx,color,subfigure}
\usepackage{amsfonts}
\usepackage{comment}
\usepackage{placeins}
\usepackage{multirow}

\usepackage{sectsty}
\usepackage{balance}
\usepackage{xspace}
\usepackage{algorithm,algorithmic}

\usepackage{appendix}
\usepackage{float}

\usepackage{epsfig,amssymb,amsmath,amsthm,amsfonts,amsbsy,mathrsfs}
\usepackage{graphicx,color}
\usepackage{amsmath}
\usepackage{amssymb}
\usepackage{epstopdf}

\usepackage{graphicx} 
\graphicspath{{figure/}}
\usepackage{lastpage}
\usepackage[format=plain,justification=raggedright,singlelinecheck=false,font=small,labelfont=bf,labelsep=space]{caption}
\usepackage{diagbox}

\allowdisplaybreaks

\usepackage[version=3]{mhchem} 
\usepackage[version=3]{mhchem}
\usepackage{subfigure}
\usepackage{placeins}
\usepackage{booktabs}

\definecolor{MatlabY}{rgb}{0.929,0.694,0.125}

\author{Yanxian Tao}
\affiliation{State Key Laboratory of Precision and Intelligent Chemistry, University of Science and Technology of China, Hefei, 230026, Anhui, China.}

\author{Lingyun Wan}
\email{wanly@ustc.edu.cn (Lingyun Wan)}
\affiliation{State Key Laboratory of Precision and Intelligent Chemistry, University of Science and Technology of China, Hefei, 230026, Anhui, China.} 

\author{Jie Liu}
\email{liujie86@ustc.edu.cn (Jie Liu)}
\affiliation{Hefei National Laboratory, University of Science and Technology of China, Hefei, 230088, China}

\title{Measurement Reduction in Orbital-Optimized Variational Quantum Eigensolver via Orbital Compression}

\begin{document}

\begin{abstract}
The variational quantum eigensolver (VQE) has emerged as one of the leading quantum algorithms for solving electronic structure problems on near-term noisy intermediate-scale quantum devices. However, its practical application to quantum chemistry remains challenging due to the limited coherence time, imperfect quantum gate fidelity, and the large number of measurements required, which together confine current electronic structure simulations to relatively small active spaces. In this work, we present an orbital-optimized VQE framework based on orbital compression, designed to improve the accuracy of electronic structure calculations while maintaining relatively small active spaces. Frozen natural orbitals (FNO) and split virtual orbitals (SVO) are first employed to construct compact active spaces for VQE simulations, leading to the FNO/SVO-VQE approach. Orbital optimization is then incorporated to further recover electron correlation effects, resulting in the FNO/SVO-OO-VQE methods. We apply the proposed method to simulate potential energy surfaces for molecular dissociation and the activation energy of formaldehyde decomposition. Numerical results demonstrate that both FNO-OO-VQE and SVO-OO-VQE improve the variational accuracy while substantially reducing measurement cost.
\end{abstract}

\section{Introduction}
Accurate solutions of electronic structure problems are central to the design of new molecules and materials in chemistry and materials science~\cite{MarFerWol21,HauJaiOne12}. Traditional wavefunction-based methods, such as configuration interaction~\cite{Eri21}, coupled-cluster theory~\cite{RevModPhys.79.291}, and multiconfigurational self-consistent field approaches~\cite{SzaMulGid12}, provide a systematic path toward improved accuracy. However, their computational cost grows rapidly with system size, thereby limiting their applicability to industrially relevant systems on classical computers~\cite{bourzac2017chemistry}. Quantum computing has therefore emerged as a promising alternative for solving electronic structure problems efficiently and accurately by harnessing quantum superposition and entanglement~\cite{daley2022practical,Pre18,cao2019quantum,McAEndAsp20,HeaFliCic21,fauseweh2024quantum,TILLY2022}.

Despite the remarkable progress in quantum computing~\cite{KimEddAna23,BluEveGei24,Gao2025Zuchongzhi3}, realizing fault-tolerant quantum computing remains a long-term goal. In the noisy intermediate-scale quantum era, the variational quantum eigensolver (VQE) has emerged as a prominent hybrid quantum-classical approach for electronic structure simulations, as it can help mitigate noise and tailor circuit depth to the available quantum resources.~\cite{peruzzo2014variational, WhiBiaAsp11, McCRomBab16, mcclean2017hybrid, romero2018strategies, ColRamDah18, MalBabKiv16}. However, due to limited coherence times and imperfect gate fidelities, applying VQE to practical electronic structure simulations involves a fundamental trade-off: one can either employ a large active space with a very shallow circuit, often at the expense of accuracy, or adopt a more expressive ansatz while restricting the active space to a size feasible on existing hardware. The former strategy is often used to demonstrate larger-scale electronic structure simulations on quantum devices~\cite{Arute2019QuantumSupremacy,JavMarHol25}. For practical applications of VQE, however, ensuring simulation accuracy is the primary consideration, which often necessitates the use of a smaller active space. In this context, for a given active space size, constructing a more suitable set of molecular orbitals is of central importance.

The construction of active spaces plays a critical role in electronic structure calculations\cite{SzaMulGid12,Keller2015,Stein2016}. An appropriately chosen active space should include the essential orbitals responsible for static and near-degeneracy correlation, such as frontier orbitals, partially occupied orbitals, and orbitals involved in bond breaking, bond formation, charge transfer, or low-lying excitations. 
In VQE-based electronic structure calculations, the choice of orbitals is even more crucial because it determines how efficiently electron correlation can be represented within a limited active space and shallow circuit\cite{McAEndAsp20,TILLY2022,liufanli22}. Canonical Hartree-Fock (HF) orbitals provide the standard starting point, but they are often suboptimal for strongly correlated systems\cite{DeGraciaTrivino2023}. Natural orbitals, especially frozen natural orbitals (FNO), offer a more compact representation by concentrating correlation into a reduced orbital space and thus lowering quantum resource requirements\cite{verma2021scaling}. A more powerful strategy is orbital-optimized VQE, in which orbital rotations are optimized together with the circuit parameters, and its state-averaged extension further enables a balanced description of multiple states in a common orbital basis\cite{mizukami2020orbital,yalouz2021state,10.1063/1.5141835,bierman2023improving,yalouz2022analytical,omiya2022analytical}. 

Although OO-VQE can substantially improve accuracy, its computational cost remains significant~\cite{mizukami2020orbital,10.1063/1.5141835}. Joint optimization of orbital rotations and circuit parameters increases both the number of optimization steps and the overall measurement overhead\cite{mizukami2020orbital,yalouz2022analytical}. In addition, each orbital update typically requires one- and two-particle reduced density matrices, making the procedure highly measurement-intensive and particularly sensitive to noise on NISQ devices\cite{yalouz2022analytical,DeGraciaTrivino2023}. To make OO-VQE practically useful, it is therefore essential to combine it with efficient orbital compression strategies that remove unnecessary degrees of freedom from the outset and thereby reduce both optimization and measurement costs\cite{verma2021scaling,kottmann2021reducing,DeGraciaTrivino2023}.

In this work, we develop a VQE framework for electronic structure calculations by combining orbital compression with explicit orbital optimization. In particular, two orbital compression schemes, frozen natural orbitals (FNO)~\cite{kumar2017frozen,pokhilko2020extension,yuan2022assessing,mochizuki2019reduction,verma2021scaling,10.1063/1.2902285} and split virtual orbitals (SVO)~\cite{shen2012coupled,kou2013hybrid}, are adopted to construct compact active spaces and subsequently integrated into the OO-VQE framework. This leads to FNO-VQE and SVO-VQE approaches that deliver higher accuracy than standard VQE without increasing the number of qubits, while the corresponding FNO-OO-VQE and SVO-OO-VQE schemes further reduce the cost of orbital optimization and the accompanying measurement overhead compared with standard OO-VQE. Benchmark calculations for representative molecular systems in the cc-pVDZ basis set show that these strategies achieve a more favorable trade-off between accuracy and quantum resource requirements, thus enhancing the practical applicability of OO-VQE to quantum chemistry in the NISQ era.

\section{Methodology}
\subsection{Variational Quantum Eigensolver}
The fermionic Hamiltonian of a many-body system in an active space is
\begin{equation}
\hat{H}_\mathrm{act} = \sum_{pq\in\mathcal{A}} h_{pq} \hat{E}_{pq} + \sum_{pqrs\in\mathcal{A}} h_{pqrs} \left\{ \hat{E}_{pq}\hat{E}_{rs} - \delta_{qr}\hat{E}_{ps} \right\}
\end{equation}
where $h_{pq}$ denotes the one-electron integrals (when an active-space Hamiltonian is employed, $h_{pq}$ should be understood as an effective one-electron term that also includes contributions arising from the frozen orbitals) and $h_{pqrs}$ denotes the two-electron integrals. $\mathcal{A}$ is the active space,  $\hat{E}_{pq} = \hat{a}^{\dagger}_{p_\alpha}\hat{a}_{q_\alpha} + \hat{a}^{\dagger}_{p_\beta}\hat{a}_{q_\beta}$ is the singlet excitation operator. Here, $\hat{a}^{\dagger}_{p_\sigma}$ and $\hat{a}_{p_\sigma}$ are the fermionic creation and annihilation operators for spatial orbital $p$ and spin $\sigma\in\{\alpha,\beta\}$. In VQE, the fermionic Hamiltonian can be mapped into a qubit Hamiltonian via Jordan-Winger~\cite{JorWig28}, or Bravyi-Kitaev~\cite{BraKit02,SeeRicLov12} transformation. After preparing the qubit Hamiltonian $\hat{H}$ and the trial wave function
\begin{equation}
    |\Psi(\boldsymbol{\theta})\rangle = \hat{U}(\boldsymbol{\theta}) |\Psi_0\rangle,
\end{equation}
the ground state energy can be determined by use of the variational principle
\begin{equation}
    E = \min_{\boldsymbol{\theta}}  \langle \Psi(\boldsymbol{\theta}) | \hat{H}_\mathrm{act} | \Psi(\boldsymbol{\theta}) \rangle,
\end{equation}
where $|\Psi_0\rangle$ is the reference state, which is taken to be the HF state. $\hat{U}(\boldsymbol{\theta})$ represents the wave function ansatz. Unitary coupled cluster (UCC) {\it ans\"atze}, e.g. UCC with single and double excitations (UCCSD), are widely used in the VQE calculations~\cite{peruzzo2014variational,romero2018strategies,lee2018generalized}. In this work, we approximate the wave function using $k$-UpCCGSD ansatz~\cite{lee2018generalized}
\begin{equation}
    |\Psi(\boldsymbol{\theta})\rangle
    =
    \prod_{l=1}^k
    \left(
    e^{\hat{T}_1^{(l)}-\hat{T}_1^{(l)\dagger}}
    e^{\hat{T}_2^{(l)}-\hat{T}_2^{(l)\dagger}}
    \right)
    |\Psi_0\rangle,
\end{equation}
where double excitations are restricted to the simultaneous promotion of an electron pair from one spatial orbital to another, so that the single- and double-excitation operators $\hat{T}_1^{(l)}$ and $\hat{T}_2^{(l)}$ in the $l$th layer is 
\begin{equation}
    \begin{split}
    &\hat{T}_1^{(l)} = \sum_{p q \alpha} t^{p_\alpha,(l)}_{q_\alpha} a^\dagger_{p_\alpha} a_{q_\alpha}, \\
    &\hat{T}_2^{(l)} = \sum_{rs\alpha\beta} t^{r_\alpha r_\beta, (l)}_{s_\alpha s_\beta} a^\dagger_{r_\alpha} a^\dagger_{r_\beta} a_{s_\alpha} a_{s_\beta}.
    \end{split}
\end{equation}
Compared with UCCSD, $k$-UpCCGSD employs a much more compact parameterization by restricting double excitations to pair excitations and introducing a k-layer product form, thereby significantly reducing circuit depth and measurement cost while retaining systematic improvability and favorable accuracy for strongly correlated systems. 

\subsection{Frozen Natural Orbital}
FNOs are a widely used orbital truncation scheme for reducing the cost of correlated electronic structure calculations. In the FNO approach, the virtual space is transformed into natural orbitals and ranked according to their occupation numbers, which reflect their contributions to electron correlation. Orbitals with negligible occupations are then frozen, yielding a compact virtual space that preserves most of the correlation energy while significantly reducing computational effort. This makes FNOs an effective and physically motivated tool for both classical and quantum electronic structure calculations.~\cite{kumar2017frozen,pokhilko2020extension,yuan2022assessing,mochizuki2019reduction,verma2021scaling}.

In this work, second-order M{\o}ller--Plesset perturbation theory (MP2) is adopted as a low-cost correlated approximation to estimate the natural occupation numbers and define the active space. In the canonical HF orbital basis, the MP2 double-excitation amplitudes are given by
\begin{equation}
t_{ij}^{ab}=\frac{\langle ij\Vert ab\rangle}{\epsilon_i+\epsilon_j-\epsilon_a-\epsilon_b},
\end{equation}
where $i, j$ denote occupied spin orbitals, $a, b$ denote virtual spin orbitals, $\langle ij\Vert ab\rangle$ is the antisymmetrized two-electron integral, and $\epsilon_p$ is the energy of corresponding spin orbital. From these amplitudes, the virtual--virtual block of the one-particle reduced density matrix (1RDM) can be constructed in the spin-orbital representation as
\begin{equation}
\gamma_{ab}^{(\mathrm{MP2})}
=
\frac{1}{2}\sum_{ijc}
t_{ij}^{ac} t_{ij}^{bc}.
\end{equation}
Then diagonalizing the virtual--virtual block $\boldsymbol{\gamma}^{(\mathrm{MP2})}_{\mathrm{vir}-\mathrm{vir}}$
\begin{equation}
\boldsymbol{\gamma}^{(\mathrm{MP2})}_{\mathrm{vir}-\mathrm{vir}}\,\mathbf{V}
=\mathbf{V}\,\mathbf{n}_{\mathrm{vir}},
\qquad
\mathbf{n}_{\mathrm{vir}}=\mathrm{diag}(n_{v
_1},n_{v_2},\ldots),
\end{equation}
yields the transformation matrix $\mathbf{V}$ for constructing the virtual natural orbital basis and the corresponding natural occupation numbers. The active virtual space can be determined either by a threshold criterion or by retaining a fixed number of orbitals:
\begin{equation}
\mathcal{V}_{\mathrm{act}}=\{v_r\in {\mathrm{vir}} \mid n_{v_r}>\tau\}
\quad \text{or} \quad
\mathcal{V}_{\mathrm{act}}=\mathrm{Top}\text{-}N_{\mathrm{vir}}^{\mathrm{act}}(\{n_v\}).
\end{equation}
In this work, the active virtual space is defined by retaining a fixed number of virtual orbitals.

\subsection{Split Virtual Orbital}
In the SVO scheme, a reduced auxiliary basis set is employed to construct a reference virtual space, and the active virtual orbitals in the larger basis are subsequently selected based on their overlap with this reference space. In this way, many high-energy virtual orbitals that make only minor contributions to correlation are excluded from the active space. Since the selection is determined by the correspondence between orbital subspaces, the number of active virtual orbitals remains unchanged across different molecular geometries. This property makes SVO especially attractive for potential energy surface calculations and reaction barrier studies~\cite{shen2012coupled,kou2013hybrid}.

The overlap matrix between the virtual orbital spaces defined in the large basis set, $\{|\psi_a\rangle\}_{a=1}^{N}$, and in the small basis set, $\{|\phi_b\rangle\}_{b=1}^{M}$, is given by
\begin{equation}
L_{ab}=\langle \psi_a | \phi_b\rangle,
\quad
a=1,\ldots,N,\quad b=1,\ldots,M.
\end{equation}
The singular value decomposition (SVD) of this overlap matrix is 
\begin{equation}
\mathbf{L}=\mathbf{V}\,\mathbf{d}\,\mathbf{U}^\dagger,
\end{equation}
where $\mathbf{V}$ is a $N \times N$ unitary matrix and $\mathbf{U}$ is a $M \times M$ unitary matrix. $\mathbf{d}$ is a $N \times M$ matrix like:
\begin{equation}
    \mathbf{d} = \begin{bmatrix}
        \widetilde{\mathbf{d}}\\
        \mathbf{0}
    \end{bmatrix}
\end{equation}
where $\widetilde{\mathbf{d}}$ is a $M \times M$ non-negative diagonal matrix whose diagonal elements $\{d_c\}$ are the singular values arranged in descending order $d_1\ge d_2\ge\cdots\ge d_M \ge0$.

The large-basis and small-basis virtual orbitals are then transformed according to
\begin{equation}
\begin{split}
|\widetilde{\psi}_c\rangle &= \sum_{a=1}^{N} |\psi_a\rangle\, V_{ac},
\qquad c=1,\ldots,N,\\
|\widetilde{\phi}_{c'}\rangle &= \sum_{b=1}^{M} |\phi_b\rangle\, U_{bc'},
\qquad c'=1,\ldots,M.
\end{split}
\end{equation}
The overlap matrix becomes diagonal
\begin{equation}
\langle \widetilde{\psi}_c | \widetilde{\phi}_{c'}\rangle
= d_{c'}\,\delta_{cc'}
\end{equation}
Thus, each rotated virtual orbital $|\widetilde{\phi}_c\rangle$ in the small basis set is uniquely associated with a rotated virtual orbital $|\widetilde{\psi}_c\rangle$ in the large basis set, and the strength of this correspondence is quantified by the singular value $d_c$. The active virtual space in the SVO method is defined as the subset of rotated large-basis virtual orbitals that has significant overlap with the reference virtual space of the small basis set.

\subsection{Orbital-Optimized Variational Quantum Eigensolver}
Orbital-Optimized Variational Quantum Eigensolver (OO-VQE) extends the standard active space VQE framework by introducing orbital rotation parameters $\boldsymbol{\kappa}$ and explicitly incorporating orbital optimization into the variational procedure. The energy is then variationally minimized with respect to both the circuit parameters $\boldsymbol{\theta}$ and orbital rotations $\boldsymbol{\kappa}$:
\begin{equation}
  E(\boldsymbol{\theta},\boldsymbol{\kappa})
  =
  \langle \Psi(\boldsymbol{\theta})| e^{-\hat{\kappa}} \hat{H}_\mathrm{act} e^{\hat{\kappa}} |\Psi(\boldsymbol{\theta})\rangle,
\end{equation}
where the orbital rotation operator is defined as $\hat{\kappa} = \sum_{pq} \kappa_{pq} \left(\hat{E}_{pq} - \hat{E}_{qp}\right)$. With fixed $\boldsymbol{\theta}$, the second order expansion of the energy function becomes
\begin{equation}
  E(\boldsymbol{\theta},\boldsymbol{\kappa})
  \approx
  E_0
  +
  \sum_{pq} g_{pq}\kappa_{pq}
  +
  \frac{1}{2}\sum_{pqrs} H_{pqrs}\kappa_{pq}\kappa_{rs},
  \label{eq:E_expand}
\end{equation}
where $E_0=\langle\Psi(\boldsymbol{\theta})|\hat{H}_\mathrm{act}|\Psi(\boldsymbol{\theta})\rangle$ is the energy in the current molecular orbital representation, $g_{pq}$ and $H_{pqrs}$ are the orbital gradient and Hessian, respectively, 
\begin{equation}
\begin{split}
g_{pq}
&=
\big\langle \Psi \big| [\hat{H},\hat{E}^{-}_{pq}] \big| \Psi \big\rangle,\\
H_{pqrs}
&=
\frac{1}{2}
\Big\langle \Psi \Big|
\big[[\hat{H},\hat{E}^{-}_{pq}],\hat{E}^{-}_{rs}\big]
+
\big[[\hat{H},\hat{E}^{-}_{rs}],\hat{E}^{-}_{pq}\big]
\Big| \Psi \Big\rangle,
\end{split}
\end{equation}
where $\hat{E}^{-}_{pq}=\hat{E}_{pq}-\hat{E}_{qp}$. By differentiating Eq.~\eqref{eq:E_expand} with respect to $\boldsymbol{\kappa}$
and setting the result to zero, one obtains the Newton--Raphson equation.
\begin{equation}
  \mathbf{H}\boldsymbol{\kappa} = -\mathbf{g}.
\label{eq:newton_raphson}
\end{equation}
The orbitals are updated iteratively by solving this equation until convergence. The Hessian $\mathbf{H}$ and gradient $\mathbf{g}$ are computed from the one- and two-electron integrals in combination with the one- and two-particle reduced density matrices (1RDM and 2RDM) obtained from VQE measurements. To lower the computational cost while preserving robust convergence behavior, we adopt the co-iterative augmented Hessian (CIAH) method implemented in PySCF.~\cite{SUN2017291,sun2017coiterativeaugmentedhessianmethod}.

\subsection{FNO/SVO-OO-VQE Workflow}
\begin{figure}[ht]
\centering
\includegraphics[width=0.45\textwidth]{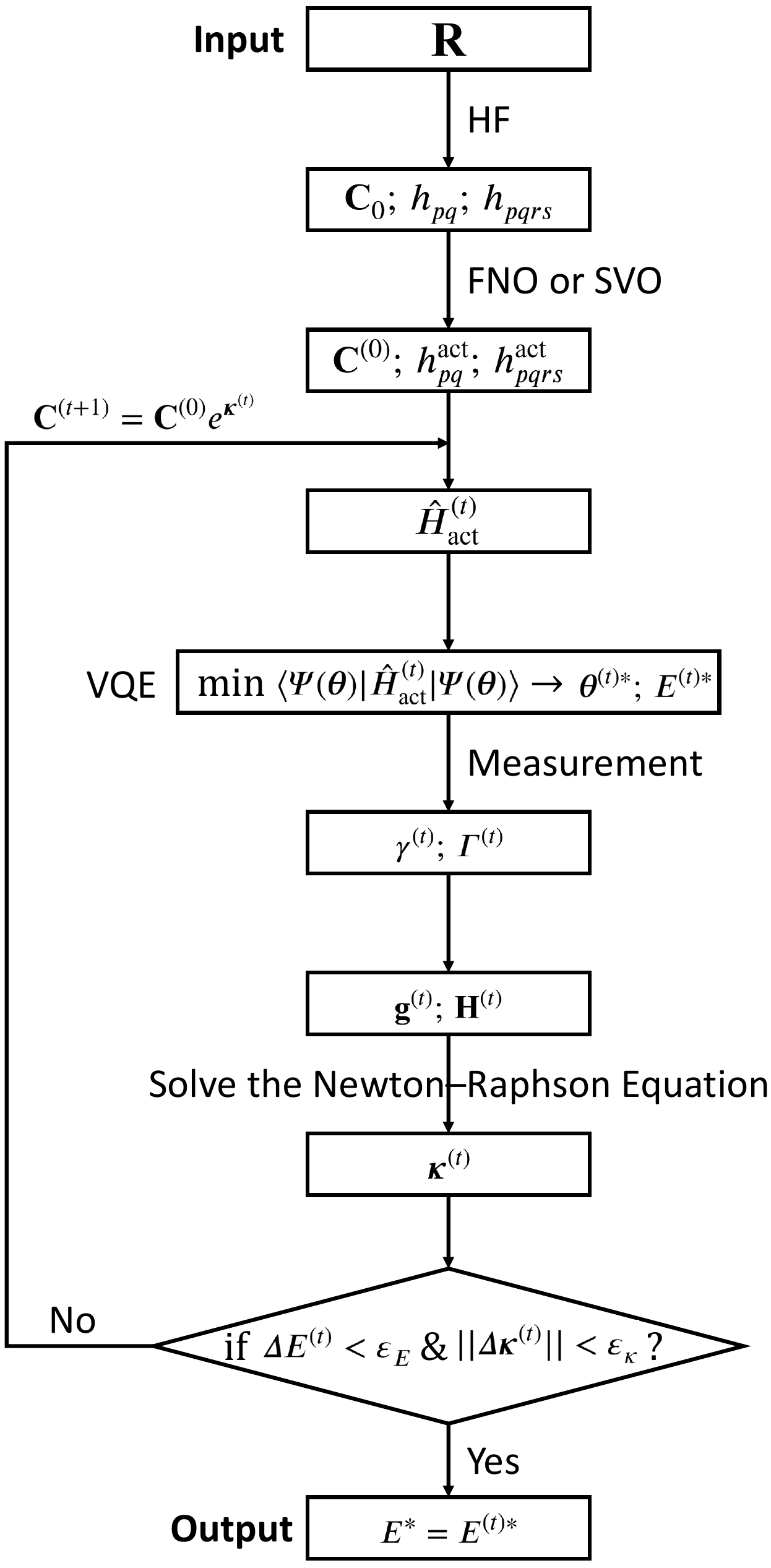}
\caption{Schematic illustration of the FNO-OO-VQE and SVO-OO-VQE workflow.}
\label{fig:fnosvooovqe}
\end{figure}

We further integrate orbital compression with orbital optimization in the VQE framework, resulting in the FNO-OO-VQE and SVO-OO-VQE approaches, for electronic structure simulations in large basis sets. In these approaches, a compact active space is first constructed using either the FNO or SVO method, thereby defining the initial orbital basis for the calculation. Following each orbital update, an inner-loop VQE calculation is carried out to optimize the circuit parameters and to evaluate the energy, as well as the 1RDM and 2RDM, which are subsequently used to update the orbitals. In this way, the circuit and orbital parameters are optimized in an alternating manner. The overall workflow is illustrated in Fig.~\ref{fig:fnosvooovqe} and summarized as follows:
\begin{enumerate}
  \item For a given molecular geometry $\mathbf{R}$, the HF calculation is performed to obtain the initial molecular orbital coefficient matrix $\mathbf{C}_0$ and the one- and two-electron integrals $h_{pq}$ and $h_{pqrs}$;
  
  \item The initial active orbitals are constructed using either FNO or SVO method;
  
  \item Constructing the active space Hamiltonian $\hat{H}_\mathrm{act}^{(t)}$, and starting the outer-loop orbital optimization:
  \begin{enumerate}
      \item For the Hamiltonian $\hat{H}_\mathrm{act}^{(t)}$, the inner-loop VQE calculation is carried out by minimizing the expectation value $\langle \Psi(\boldsymbol{\theta}) | \hat{H}_\mathrm{act}^{(t)} | \Psi(\boldsymbol{\theta}) \rangle$, yielding the optimized circuit parameters $\boldsymbol{\theta}^{(t)*}$ and the corresponding energy $E^{(t)*}$;
      
      \item The 1RDM $\gamma^{(t)}$ and 2RDM $\Gamma^{(t)}$ are evaluated from the optimized VQE wavefunction;
      
      \item The orbital gradient $\mathbf{g}^{(t)}$ and Hessian $\mathbf{H}^{(t)}$ are computed from $\gamma^{(t)}$ and $\Gamma^{(t)}$;
      
      \item The orbital rotation parameters $\boldsymbol{\kappa}^{(t)}$ are obtained by solving the Newton--Raphson equation Eq.~\ref{eq:newton_raphson};
      
      \item The orbitals are updated according to $\mathbf{C}^{(t+1)} = \mathbf{C}^{(t)} e^{\boldsymbol{\kappa}^{(t)}}$ and the above steps are repeated.
  \end{enumerate}
  
  \item When the energy difference $\Delta E^{(t)}$ between two consecutive iterations is smaller than a threshold $\varepsilon_E$, and the norm of the orbital update $||\Delta \boldsymbol{\kappa}^{(t)}||$ is smaller than a threshold $ \varepsilon_\kappa$, the orbital optimization is regarded as converged. And the final variational energy is taken as $E^* = E^{(t)*}$.
\end{enumerate}

\section{Results and Discussion}
\subsection{Computational Details}
We use PySCF~\cite{sun2018pyscf} to compute one- and two-electron integrals and perform the orbital optimization. The Jordan--Wigner transformation is carried out using OpenFermion~\cite{mcclean2020openfermion}. The optimization of the variational parameters is performed with SciPy~\cite{virtanen2020scipy}. The VQE wavefunction is parameterized by the $k$-UpCCGSD ansatz~\cite{lee2018generalized}.

We use $(m,n)$ to denote the active space, where $m$ and $n$ are the number of active electrons and molecular orbitals, respectively. All calculations are performed with the cc-pVDZ basis set. In the SVO scheme, the cc-pVDZ basis set is used as the larger one and the small auxiliary basis sets are summarized in Table~\ref{tab:settings}. In addition, the number of qubits and the number of parameters in the quantum circuits, are also listed in Table~\ref{tab:settings}.

\begin{table*}
  \centering
  \caption{Molecular systems and computational parameters. The second column shows the total number of electrons and molecular orbitals in the form $(m,n)$ at the level of the cc-pVDZ basis set. The third column lists the small auxiliary basis sets used in the SVO scheme. $N_\mathrm{qubit}$ and $N_\theta$ denote the number of qubits and parameters in the quantum circuit, respectively.}
  \label{tab:settings}
  \scalebox{1.0}{
  \begin{tabular}{c cccccc}
  \toprule
  Molecule & cc-pVDZ & Small Basis Set & Active Space & $N_\mathrm{qubit}$ & Ansatz & $N_\theta$ \\
  \midrule
  \ce{LiH}  & $(4,19)$  & STO-6G & $(4,6)$  & 12 & $3$-UpCCGSD & 90 \\
  \ce{H2O}  & $(10,24)$ & Minao  & $(10,7)$ & 14 & $3$-UpCCGSD & 126 \\
  \ce{N2}   & $(10,26)$ & Minao  & $(10,8)$ & 16 & $6$-UpCCGSD & 336 \\
  \ce{H2CO} & $(16,38)$ & Minao  & $(8,8)$  & 16 & $6$-UpCCGSD & 336 \\
  \bottomrule
  \end{tabular}
  }
\end{table*}

\subsection{Benchmark Tests}
We first employ the \ce{LiH} molecule as a benchmark system to systematically assess the accuracy and efficiency of FNO-VQE, SVO-VQE, FNO-OO-VQE, and SVO-OO-VQE. All calculations are performed using the 
$3$-UpCCGSD ansatz, which offers a practical compromise between computational cost and accuracy. It is therefore well suited for evaluating the performance of the FNO and SVO orbital compression schemes. For the SVO scheme, STO-6G is employed as the small auxiliary basis set, yielding an active space of $(4,6)$. To enable a direct comparison among the various methods, all VQE calculations in this section are carried out using active spaces of the same size, with the numbers of qubits and variational parameters kept identical. 

\begin{figure*}[ht]
\centering
\includegraphics[width=1.0\textwidth]{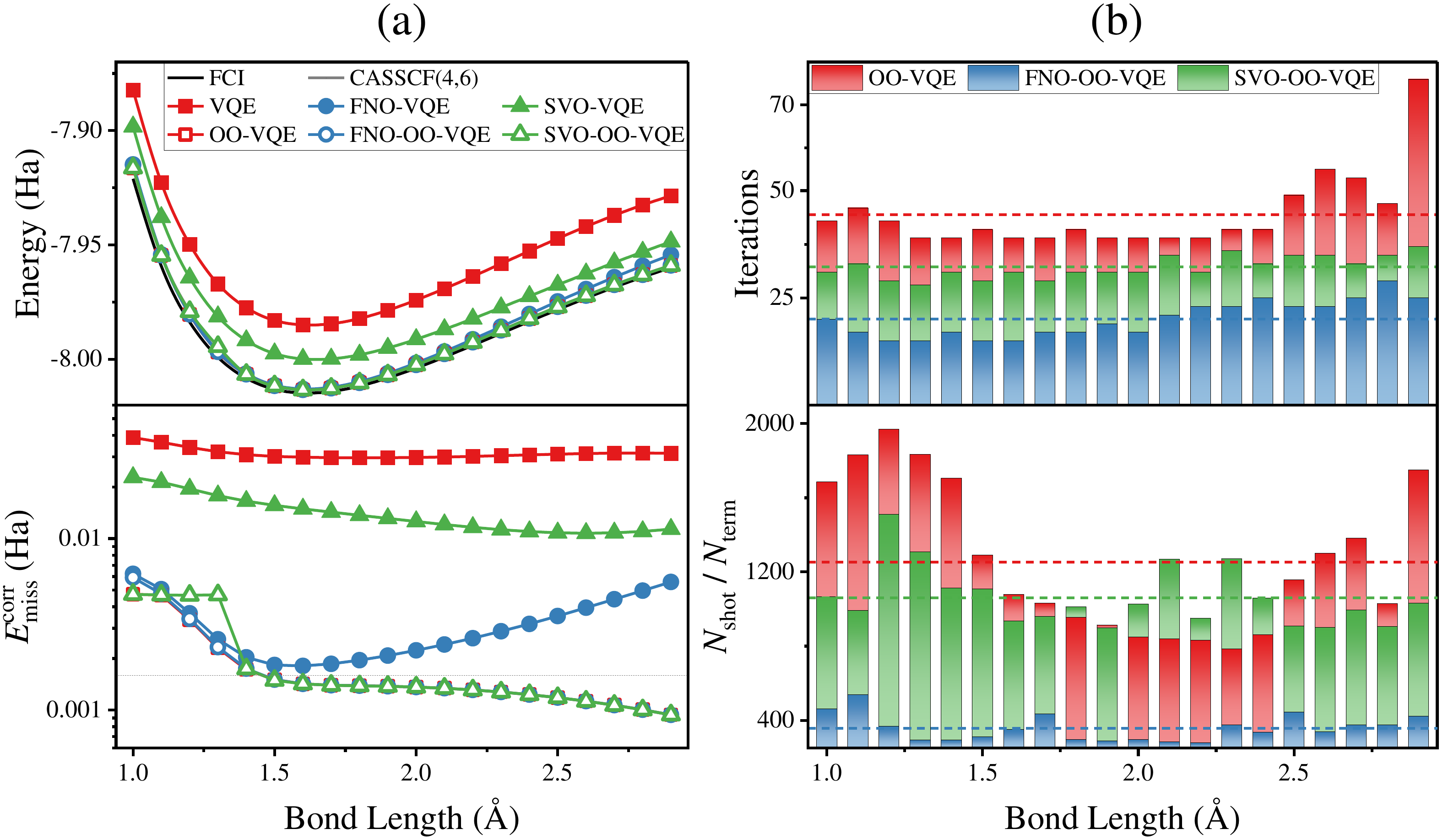}
\caption{(a) Potential energy surfaces of \ce{LiH} and the corresponding errors in correlation energy ($E_\mathrm{miss}^\mathrm{corr}$) with respect to the exact diagonalization results (labeled as ``FCI''). All VQE calculations use the same active space $(4,6)$. (b) Number of iterations in orbital optimization for different OO-VQE methods and the ratio of the total number of measurements to the number of qubit Hamiltonian terms $N_\mathrm{shot}/N_\mathrm{term}$. Dashed lines denote the averages over all \ce{LiH} geometries for each method.}
\label{fig:lihresult}
\end{figure*}

Figure~\ref{fig:lihresult}(a) shows the \ce{LiH} potential energy surfaces and the corresponding errors in correlation energy, $E_\mathrm{miss}^\mathrm{corr}$, with respect to the exact diagonalization results, labeled as ``FCI''. Compared with standard VQE based on canonical Hartree–Fock orbitals, both FNO-VQE and SVO-VQE produce substantially lower energies throughout the full dissociation curve. FNO-VQE shows the best overall performance, with $E_\mathrm{miss}^\mathrm{corr}$ of only about 0.003 Hartree relative to FCI. These results demonstrate that appropriate orbital compression can effectively identify the most relevant orbitals and define a compact active space that achieves significantly improved accuracy without increasing the number of qubits or the circuit depth.

Furthermore, after incorporating orbital optimization into these VQE methods, the potential energy surfaces obtained from all OO-VQE variants are in good agreement with the complete active space self-consistent field (CASSCF) results. Moreover, their $E_\mathrm{miss}^\mathrm{corr}$ values relative to FCI remain within chemical accuracy near the equilibrium geometry and along the bond dissociation coordinate, indicating that explicit orbital optimization can recover part of the missing correlation energy and further lower the variational energy. Despite their comparable accuracy, however, the computational costs of the different OO-VQE approaches differ substantially. As shown in the upper panel of Fig.~\ref{fig:lihresult}(b), FNO-OO-VQE and SVO-OO-VQE require fewer orbital optimization iterations than standard OO-VQE, reducing the iteration count to $45.2\%$ and $75.5\%$ of that of OO-VQE, respectively. This reduction reflects the improved quality of the initial orbitals provided by orbital compression and directly translates into fewer repeated VQE evaluations.

Because each orbital optimization loop requires evaluation of 1RDM, 2RDM and the energy, the improved accuracy of OO-VQE comes at the expense of a substantial increase in measurement cost. Since all methods considered here have the same, and relatively large, number of qubit Hamiltonian terms $N_\mathrm{term}$, we adopt the ratio $N_\mathrm{shots}/N_\mathrm{term}$ in Fig.~\ref{fig:lihresult}(b) as a compact metric for the measurement cost. Both FNO-OO-VQE and SVO-OO-VQE show a clear reduction in measurement cost compared with standard OO-VQE, with FNO-OO-VQE reducing the total cost to only $28.5\%$ of that of OO-VQE. Importantly, this reduction is significantly larger than the decrease in the number of orbital-optimization iterations alone, indicating that FNO not only improves the initial orbitals and lowers the number of outer-loop iterations, but also increases the efficiency of the inner-loop VQE parameter optimization performed within each orbital-update cycle. As a result, the overall computational efficiency is substantially enhanced.

\subsection{Molecular Potential Energy Surfaces}

\begin{figure*}[ht]
\centering
\includegraphics[width=1.0\textwidth]{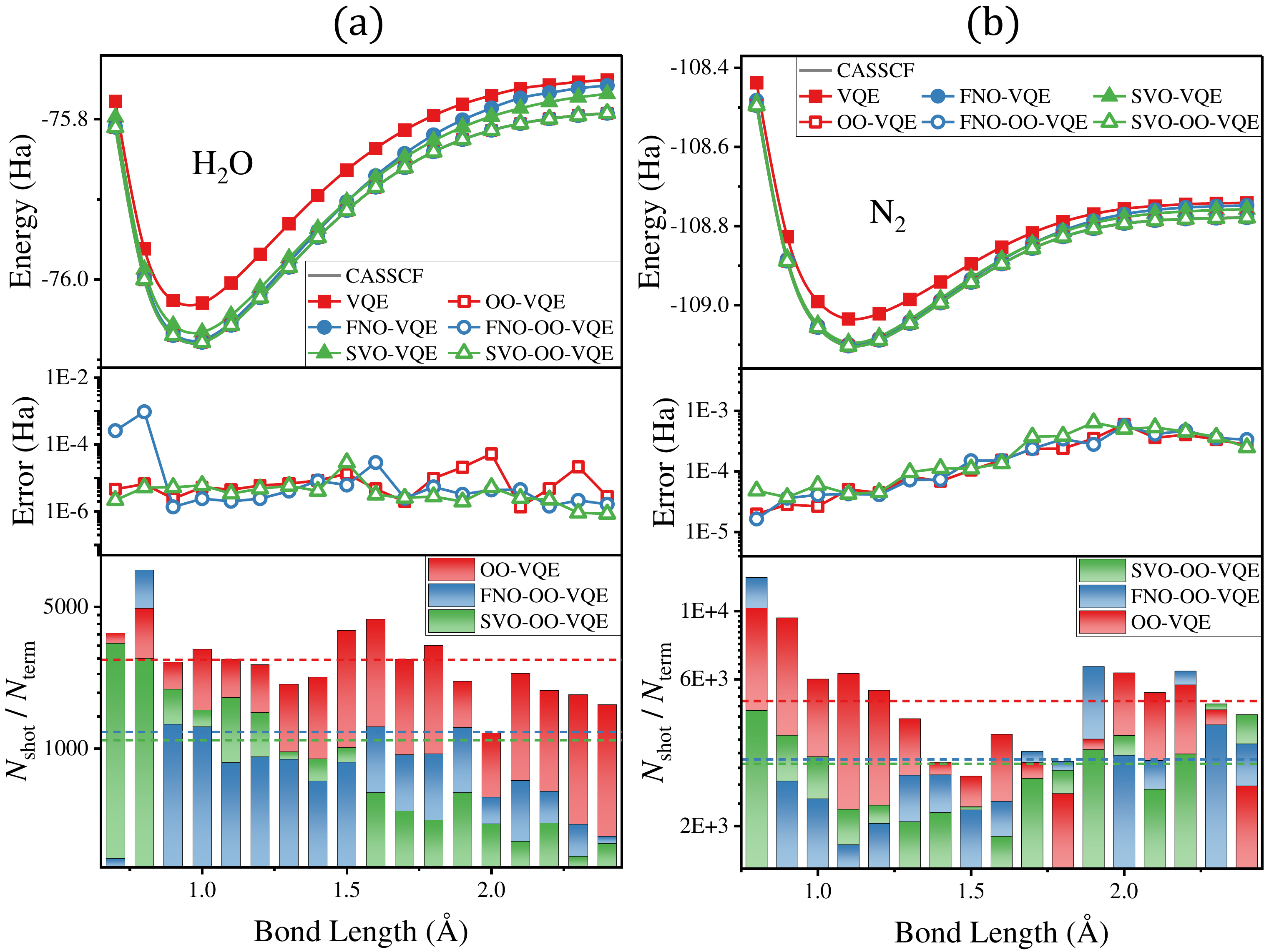}
\caption{Potential energy surfaces of \ce{H2O} (a) and \ce{N2} (b) computed with different VQE methods, together with the corresponding energy errors relative to CASSCF, and the ratio $N_\mathrm{shot}/N_\mathrm{term}$ used to characterize the measurement cost. The active spaces for \ce{H2O} and \ce{N2} are $(10,7)$ and $(10,8)$, respectively. Dashed lines in the measurement cost plots denote the averages of the corresponding methods over all geometries considered for each molecule.}
\label{fig:moleresult}
\end{figure*}

In the following, we apply the OO-VQE methods to more challenging molecular systems, namely \ce{H2O} and \ce{N2}, for potential energy surface calculations. These systems constitute more demanding benchmarks for current quantum algorithms. The computational details are summarized in Table~\ref{tab:settings}, and the corresponding results are shown in Fig.~\ref{fig:moleresult}. Consistent with the trends observed for \ce{LiH}, the energies obtained with FNO-VQE and SVO-VQE are significantly lower than those from standard VQE, indicating that both FNO and SVO provide improved initial orbitals. Consequently, FNO-OO-VQE and SVO-OO-VQE substantially reduce the measurement cost relative to OO-VQE while maintaining an accuracy comparable to that of CASSCF. Specifically, for \ce{H2O}, the total measurement cost is reduced to $44.1\%$ and $40.1\%$ of that of OO-VQE by FNO-OO-VQE and SVO-OO-VQE, respectively, whereas for \ce{N2} the corresponding values are $64.7\%$ and $62.5\%$. These results demonstrate that the combination of orbital compression and orbital optimization remains effective for medium-sized molecular systems.

Overall, FNO-OO-VQE and SVO-OO-VQE exhibit very similar performance for both molecules considered here. Their nearly identical accuracy suggests that orbital optimization ultimately converges to similar optimal orbital subspaces, whereas their similar computational costs indicate that the two compression schemes retain a comparable number of effective degrees of freedom. Nevertheless, the two approaches offer distinct practical advantages. FNO allows flexible tuning of the active space through either an occupation threshold or a fixed number of retained virtual orbitals, making it convenient for balancing accuracy against computational cost. In contrast, SVO defines the active space through the projection between the virtual spaces of large and small basis sets and therefore often provides better continuity and numerical stability along a potential energy surface scan. In practice, the choice between these two orbital compression schemes may be made according to the specific requirements of the target problem.

Although the absolute energies obtained from FNO-OO-VQE and SVO-OO-VQE still deviate from the exact reference values from FCI or DMRG because of the relatively small active space and the approximate ansatz, the structural and thermochemical quantities derived from the resulting potential energy surfaces remain highly reliable. Table~\ref{tab:bondandenergy} summarizes the equilibrium bond lengths and bond dissociation energies extracted from the potential energy surfaces shown in Fig.~\ref{fig:moleresult}. For all molecules considered, the errors in the equilibrium bond lengths predicted by FNO-OO-VQE and SVO-OO-VQE are only about $0.002$~\AA. The corresponding errors in the bond dissociation energies are on the order of millihartree for all systems except \ce{H2O}, for which the deviation is $35$~millihartree. This deviation mainly orginates from the missing of the dynamic correlation energy, which can be recovered by NEVPT2 (second-order n-electron valence state perturbation theory) if necessary, reducing the error to $3.68$~millihartree. Relative to the deviations in the absolute energies, the errors in these differential quantities are negligible, indicating that the computed potential energy surfaces remain nearly parallel to the exact reference curves over a broad range of geometries and therefore reproduce energetic trends accurately. This behavior suggests that the residual errors in FNO-OO-VQE and SVO-OO-VQE are dominated by an approximately geometry-independent systematic shift, which enables reasonably accurate predictions of structural and thermochemical differences. Such a cancellation, however, may become less effective when the missing correlation energy varies significantly with geometry. In these cases, more robust accuracy may require a larger active space, a more expressive ansatz, or further improvements in the orbital optimization procedure.

\begin{table}
  \centering
  \caption{Equilibrium bond lengths (a) and bond dissociation energies (b) of molecules calculated with the FNO-OO-VQE and SVO-OO-VQE methods. For \ce{LiH}, the FCI results are used as the reference, whereas for \ce{H2O} and \ce{N2}, the reference values are taken from density matrix renormalization group (DMRG) calculations.}
  \label{tab:bondandenergy}
  \subtable[Equilibrium Bond Length / \AA]{
  \scalebox{1.0}{
  \begin{tabular}{c ccc}
  \toprule
  Molecule & Reference & CASSCF & FNO/SVO-OO-VQE \\
  \midrule
  \ce{LiH}  & 1.615 & 1.617 & 1.617 \\
  \ce{H2O}  & 0.965 & 0.969 & 0.968 \\
  \ce{N2}   & 1.120 & 1.116 & 1.117 \\
  \bottomrule
  \end{tabular}
  }
  }
  \subtable[Bond Dissociation Energy / millihartree]{
  \scalebox{1.0}{
  \begin{tabular}{c ccc}
  \toprule
  Molecule & Reference & CASSCF & FNO/SVO-OO-VQE \\
  \midrule
  \ce{LiH}  & 54.89 & 54.42 & 54.42 \\
  \ce{H2O}  & 330.3 & 294.7 & 294.9 \\
  \ce{N2}   & 313.8 & 323.4 & 323.7 \\
  \bottomrule
  \end{tabular}
  }
  }
\end{table}

\subsection{Reaction Pathway Simulations}
\begin{figure}[ht]
\centering
\includegraphics[width=0.8\columnwidth]{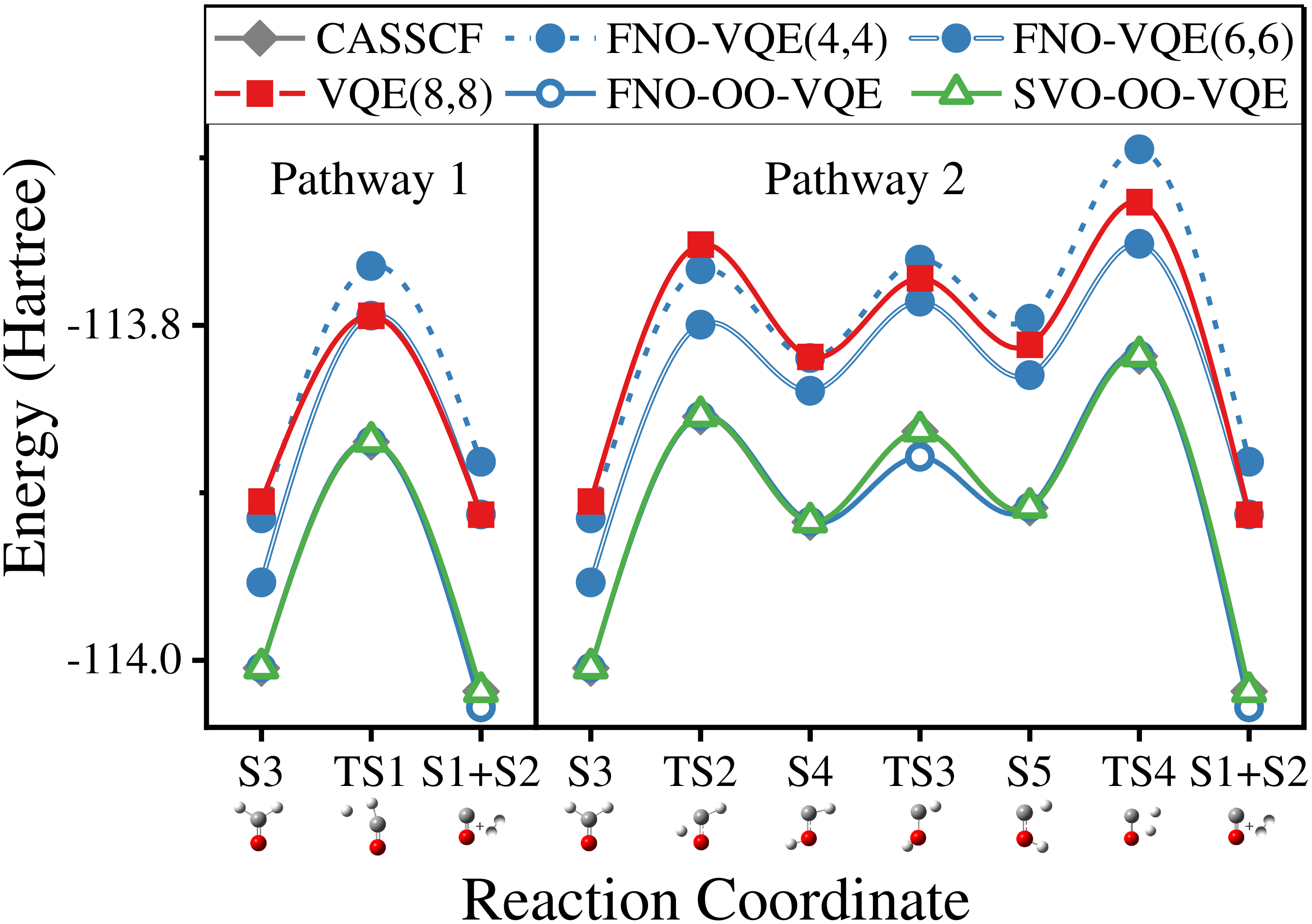}
\caption{Energy profiles for formaldehyde decomposition reaction. The left and right panels show the two reaction pathways, respectively. $\mathrm{S3}$ denotes the reactant \ce{H2CO}, $\mathrm{S1+S2}$ denotes the products \ce{H2} and \ce{CO}, $\mathrm{TS}$ denotes the transition state, and $\mathrm{S4}$ and $\mathrm{S5}$ are intermediates in pathway 2. Unless otherwise specified in the legend, the active space is $(8,8)$.}
\label{fig:h2coresult}
\end{figure}

Finally, we apply the OO-VQE methods to study chemical reaction mechanisms, where activation energies are determined by relative energies between different molecular geometries. Provided that the absolute energy error varies smoothly along the reaction coordinate, or is largely canceled in energy differences, activation barriers can still be predicted reliably. Here, we use FNO-OO-VQE and SVO-OO-VQE to simulate the decomposition pathways of \ce{H2CO}. The computational details are given in Table~\ref{tab:settings}, and the corresponding results are shown in Fig.~\ref{fig:h2coresult}. The energy profiles obtained with both methods are in excellent agreement with the CASSCF results in the same active space, indicating that they can reliably capture the relative energy changes along the reaction coordinate. Because pathway 1 is the dominant decomposition channel of \ce{H2CO}, we further evaluate its activation energy, as summarized in Table~\ref{tab:h2coactivation}. The activation energies computed by FNO-OO-VQE and SVO-OO-VQE differ from the CASSCF value by only $0.042$ and $0.115$ kcal/mol, respectively, and are also consistent with previously reported theoretical results~\cite{GULVI2024114605, wang2017new, FellerBarrier2000}. These results confirm the validity and practical promise of FNO-OO-VQE and SVO-OO-VQE for reaction pathway simulations and activation energy calculations.

\begin{table}
  \centering
  \caption{Activation energies (in kcal/mol) of pathway 1 for formaldehyde decomposition. The first column reports the range of activation energies obtained from previous theoretical calculations~\cite{GULVI2024114605, wang2017new, FellerBarrier2000}}
  \label{tab:h2coactivation}
  \begin{tabular}{cccc}
  \toprule
  Reference & CASSCF & FNO-OO-VQE & SVO-OO-VQE \\
  \midrule
  81.40$\sim$85.94 & 84.79 & 84.82 & 84.90      \\
  \bottomrule
  \end{tabular}
\end{table}

Figure~\ref{fig:h2coresult} also compares the energy accuracy of FNO-VQE and standard VQE for different active-space sizes. Along all decomposition pathways of \ce{H2CO}, FNO-VQE with a $(6,6)$ active space consistently outperforms standard VQE with a larger $(8,8)$ active space. Furthermore, at some molecular geometries, FNO-VQE with only a $(4,4)$ active space already reaches the accuracy obtained by standard VQE in the $(8,8)$ active space. Since a smaller active space directly implies fewer qubits, shallower circuits, and significantly lower measurement cost, these results show that a properly designed FNO-based compact active space can simultaneously reduce quantum resource requirements and improve variational accuracy, thus offering a more favorable accuracy-to-cost trade-off.

\section{Conclusion}

In this work, we developed the FNO-OO-VQE and SVO-OO-VQE methods for quantum chemical calculations in large basis sets by combining orbital compression with explicit orbital optimization. The present results show that this combination provides a practical route for improving the balance between accuracy and computational cost in VQE-based simulations. By constructing compact active spaces through the FNO or SVO schemes and subsequently refining the orbitals variationally, the proposed methods retain much of the accuracy gain associated with OO-VQE while substantially reducing the corresponding measurement overhead. For the molecular systems considered here, this strategy yields accurate potential energy surfaces and reliable structural, thermochemical, and reaction-profile quantities at a significantly lower cost than standard OO-VQE.

It is worth mentioning that the performance of the FNO-OO-VQE and SVO-OO-VQE methods remains sensitive to the compressed active space. Although orbital compression is designed to remove orbitals that contribute only weakly to correlation, the importance of individual orbitals may vary with molecular geometry, bonding pattern, and electronic state. As a result, the truncation error introduced at the orbital-compression stage may not be uniform along a potential energy surface, which can reduce the degree of error cancellation in differential quantities such as bond dissociation energies and activation barriers. Therefore, it is necessary to integrate these methods with perturbation theory to further improve the accuracy~\cite{RyaIZmGen21,TamGalRic23,LiuLiYan24}. 

In addition, while orbital compression lowers the cost of orbital optimization, the overall OO-VQE procedure remains measurement intensive because each orbital update still requires repeated evaluations of the energy as well as the 1RDM and 2RDM. This issue is particularly relevant for implementations on noisy quantum hardware, where measurement noise and finite sampling errors may directly affect the stability of the orbital optimization. Therefore, it is necessary to  
integrate the present framework with improved measurement-reduction techniques, more robust reduced-density-matrix estimators, and error-mitigation strategies to further reduce the practical cost of OO-VQE on quantum hardware~\cite{Rubin2018,Zhao2021,Peng2023}. 

Overall, the present work shows that orbital compression and orbital optimization can act synergistically in VQE-based quantum chemistry. Although substantial challenges remain before such methods can be routinely applied on quantum hardware, the results reported here indicate that chemically motivated orbital design provides an effective route toward more accurate and resource-efficient quantum simulations beyond minimal basis sets.

\section{Acknowledgements}
This work is supported by Innovation Program for Quantum Science and Technology (2021ZD0303306), the National Natural Science Foundation of China (22422304, 22073086, 21825302, 22288201, 22393913 and 22303090), the Strategic Priority Research Program (XDB0450101) and the robotic AI-Scientist platform of the Chinese Academy of Sciences, Anhui Initiative in Quantum Information Technologies (AHY090400).

\footnotesize{
\bibliography{ref}
}


\end{document}